\documentclass[preprint,showpacs,showkeys,preprintnumbers,amsmath,amssymb]{revtex4}
 \usepackage{dcolumn}
 \usepackage{bm}

\begin{document}

 \title{ Hawking radiation of Dirac particles via tunneling from Kerr black hole }

 \author{Ran Li}
 \thanks{Electronic mail: liran05@lzu.cn}
 \author{Ji-Rong Ren}
 \thanks{Corresponding author. Electronic mail:renjr@lzu.edu.cn}
 \author{Shao-Wen Wei}
 \affiliation{Institute of Theoretical Physics, Lanzhou University, Lanzhou, 730000, Gansu, China}

 \begin{abstract}

 We investigated Dirac Particles' Hawking radiation
 from event horizon of Kerr black hole in terms of the tunneling
 formalism. Applying WKB approximation to the general covariant Dirac
 equation in Kerr spacetime background,
 we obtain the tunneling probability for fermions and Hawking temperature of
 Kerr black hole. The result obtained by taking the fermion
 tunneling into account is consistent with the previous literatures.

 \end{abstract}



 \maketitle

 Hawking\cite{hawking} discovered the thermal radiation of a collapsing black hole
 using the techniques of quantum field theory in curved
 spacetime. Since the Hawking radiation relates the
 theory of general relativity with quantum
 field theory and statistical thermodynamics, it is generally
 believed that a deeper understanding of Hawking radiation may shed
 some lights on seeking the underlying quantum gravity.
 Since then, several derivations of Hawking radiation have been
 proposed. The original method presented by Hawking is direct but
 complicated to be generalized to other spacetime backgrounds. In
 recent years, a semi-classical derivation of Hawking radiation as
 a tunneling process\cite{parikh} has been developed and has
 already attracted a lot of
 attention. In this method, the imaginary part of the action is
 calculated using the null geodesic equation. Zhang and Zhao extended this method to
 Reissner-Nordstr\"{o}m black hole\cite{zhangjhep} and Kerr-Newman black hole\cite{zhangplb}.
 M. Angheben \textit{et al} \cite{angheben} also proposed a
 derivation of Hawking radiation by calculating the particles'
 classical action from the Hamilton-Jacobi equation, which is an
 extension of the complex path analysis of T. Padmanabhan \textit{et
 al} \cite{padmanabhan}. All of these approaches to tunneling
 used the fact that the tunneling probability for the classically
 forbidden trajectory from inside to outside the horizon is given
 by
 \begin{eqnarray}
 \Gamma=\textrm{exp}(-\frac{2}{\hbar}\textrm{Im}I)\;.
 \end{eqnarray}
 where $I$ is the classical action of the trajectory.
 The crucial thing in tunneling formalism is to calculate the
 imaginary part of classical action. The difference between these
 two methods consists in how the classical action is calculated.
 For a detailed comparison of the Hamilton-Jacobi method and the
 Null-Geodesic method, one can see \cite{kerner}.

 However, most of authors only considered the scalar particles'
 radiation. Very recently, a new calculation concerning fermions'
 radiation from the stationary spherical symmetric black hole was
 done by R. Kerner and R. B. Mann in \cite{kernerarxiv}. This
 method have been generalized to BTZ black hole by us in \cite{liplb} and
 dynamical black hole in \cite{vanzo}.
 In this letter, we will generalize the tunneling method presented in \cite{kernerarxiv}
 to calculate the Dirac particles' Hawking radiation from Kerr black hole.
 Starting with the general covariant Dirac
 equation in curved background, we calculate the tunneling
 probability and Hawking temperature by using WKB approximation.
 The result obtained by taking the fermion
 tunneling into account is consistent with the previous literatures.

 The Kerr black hole solution in Boyer-Linquist coordinates is given by
 \begin{eqnarray}
 ds^2=&-&\frac{\Delta-a^2\textrm{sin}^2\theta}{\Sigma}dt^2
 -2a\textrm{sin}^2\theta\cdot\frac{r^2+a^2-\Delta}{\Sigma}dtd\phi\nonumber\\
 &+&\frac{(r^2+a^2)^2-\Delta
 a^2\textrm{sin}^2\theta}{\Sigma}\cdot\textrm{sin}^2\theta d\phi^2
 +\frac{\Sigma}{\Delta}dr^2+\Sigma d\theta^2\;\;,
 \end{eqnarray}
 where
 \begin{eqnarray*}
 \Sigma&=&r^2+a^2\textrm{cos}^2\theta\;\;,\\
 \Delta&=&r^2-2Mr+a^2=(r-r_+)(r-r_-)\;\;,\\
 r_\pm&=&M\pm\sqrt{M^2-a^2}\;\;.
 \end{eqnarray*}
 We have assumed the non-extremal condition $M>a$, so that $r_+$ and $r_-$
 correspond to the outer event horizon and the inner event horizon
 respectively. The determinant of the metric is
 \begin{eqnarray*}
 \sqrt{-g}=\Sigma\;\textrm{sin}\theta\;\;,
 \end{eqnarray*}
 and the inverse of the metric of $(t, \phi)$ parts is
 \begin{eqnarray*}
 g^{tt}&=&-\frac{(r^2+a^2)^2-\Delta
 a^2\textrm{sin}^2\theta}{\Sigma\Delta}\;\;,\\
 g^{\phi\phi}&=&\frac{\Delta-a^2\textrm{sin}^2\theta}
 {\Sigma\Delta\textrm{sin}^2\theta}\;\;,\\
 g^{t\phi}&=&-\frac{a(r^2+a^2-\Delta)}{\Sigma\Delta}\;\;.
 \end{eqnarray*}

 Now we calculate the Dirac particles' Hawking radiation from
 the outer event horizon of Kerr
 black hole via tunneling formulism. For simplicity,
 we only consider the massless spinor field $\Psi$ obeys the
 general covariant Dirac equation
 \begin{eqnarray}
 -i\hbar\gamma^ae_a^\mu\nabla_\mu\Psi=0\;,
 \end{eqnarray}
 where $\nabla_\mu$ is the spinor covariant derivative defined by
 $\nabla_\mu=\partial_\mu+\frac{1}{4}\omega_\mu^{ab}\gamma_{[a}\gamma_{b]}$,
 and $\omega_\mu^{ab}$ is the spin connection, which can be given
 in terms of the tetrad $e_a^\mu$.
 The $\gamma$ matrices are selected as
 \begin{eqnarray*}
 \gamma^0&=&\left(%
 \begin{array}{cc}
  i & 0 \\
  0 & -i \\
 \end{array}%
 \right)\;\;,\\
 \gamma^1&=&\left(%
 \begin{array}{cc}
  0 & \sigma^3 \\
  \sigma^3 & 0 \\
 \end{array}%
 \right)\;\;,\\
 \gamma^2&=&\left(%
 \begin{array}{cc}
  0 & \sigma^1 \\
  \sigma^1 & 0 \\
 \end{array}%
 \right)\;\;,\\
 \gamma^3&=&\left(%
 \begin{array}{cc}
  0 & \sigma^2 \\
  \sigma^2 & 0 \\
 \end{array}%
 \right)\;\;,
 \end{eqnarray*}
 where the matrices $\sigma^k(k=1,2,3)$ are the Pauli matrices.
 According to the line element (2), the tetrad fields $e_a^\mu$ can
 be selected to be
 \begin{eqnarray*}
 e_0^\mu&=&(\sqrt{-g^{tt}}, 0, 0, \frac{-g^{t\phi}}{\sqrt{-g^{tt}}})\;,\\
 e_1^\mu&=&(0, \sqrt{\frac{\Delta}{\Sigma}}, 0, 0)\;,\\
 e_2^\mu&=&(0, 0, \frac{1}{\sqrt{\Sigma}}, 0)\;,\\
 e_3^\mu&=&(0, 0, 0, \frac{1}{\sqrt{g_{\phi\phi}}})\;.
 \end{eqnarray*}

 We employ the ansatz for the spin-up spinor field $\Psi$ as
 following\cite{kernerarxiv,vanzo}
 \begin{eqnarray}
 \Psi=\left(%
 \begin{array}{c}
  A(t,r,\theta,\phi) \\
  0\\
  B(t,r,\theta,\phi) \\
  0
 \end{array}%
 \right)\textrm{exp}\big[\frac{i}{\hbar}I(t,r,\theta,\phi)\big]\;.
 \end{eqnarray}
 Note that we will only analysis the spin-up case since the
 spin-down case is just analogous.
 In order to apply WKB approximation, we can insert the ansatz
 for spinor field $\Psi$ into the general covariant Dirac equation. Dividing by the
 exponential term and neglecting the terms with $\hbar$, one can
 arrive at the following four equations
 \begin{eqnarray}
 \left\{
  \begin{array}{ll}
  iA(\sqrt{-g^{tt}}\partial_t I-\frac{g^{t\phi}}{\sqrt{-g^{tt}}}\partial_\phi I)
  +B\sqrt{\frac{\Delta}{\Sigma}}\partial_r I =0  \;,  \\
  (\frac{1}{\sqrt{\Sigma}}\partial_\theta I
  +i\frac{1}{\sqrt{g_{\phi\phi}}}\partial_\phi I)B=0  \;,  \\
  A\sqrt{\frac{\Delta}{\Sigma}}\partial_r I
  -iB(\sqrt{-g^{tt}}\partial_t I-\frac{g^{t\phi}}{\sqrt{-g^{tt}}}\partial_\phi I)=0  \;,  \\
  (\frac{1}{\sqrt{\Sigma}}\partial_\theta I
  +i\frac{1}{\sqrt{g_{\phi\phi}}}\partial_\phi I)A=0  \;.
  \end{array}
 \right.
 \end{eqnarray}
 Note that although $A$ and $B$ are not constant, their
 derivatives and the components $\omega_\mu$ are all of the factor
 $\hbar$, so can be neglected to the lowest order in WKB
 approximation. Because we only consider the Dirac field outside
 the event horizon, the condition $\Delta\geq 0$ is always satisfied in
 the above equations. The second and fourth equations indicate that
 \begin{eqnarray}
 \frac{1}{\sqrt{\Sigma}}\partial_\theta I
  +i\frac{1}{\sqrt{g_{\phi\phi}}}\partial_\phi I=0\;\;.
 \end{eqnarray}
 From the first and third equations one can see that
 these two equations have a non-trivial solution for $A$ and $B$ if and only if the
 determinant of the coefficient matrix vanishes. Then we can get
 \begin{eqnarray}
 (\sqrt{-g^{tt}}\partial_t I-\frac{g^{t\phi}}{\sqrt{-g^{tt}}}\partial_\phi I)^2
 -\frac{\Delta}{\Sigma}(\partial_r I)^2=0\;.
 \end{eqnarray}
 Because there are two killing vectors $(\partial/\partial
 t)^{\mu}$ and $(\partial/\partial\phi)^{\mu}$ in Kerr-Newman
 spacetime, we can separate the variables for $I(t,r,\theta,\phi)$ as
 following
 \begin{eqnarray}
 I=-\omega t+j\phi+R(r,\theta)+K\;,
 \end{eqnarray}
 where $\omega$ and $j$ are Dirac particle's energy and angular
 momentum respectively and $K$ is a complex constant. Inserting
 it to equation (7), one can arrive
 \begin{eqnarray}
 (\sqrt{-g^{tt}}\omega+\frac{g^{t\phi}}{\sqrt{-g^{tt}}}j)^2
 -\frac{\Delta}{\Sigma}(\partial_r R)^2=0\;.
 \end{eqnarray}
 Equation (6) indicates that $R(r,\theta)$ is a complex function.
 Now solving the equation (9) for definite $\theta_0$\ yields
 \cite{angheben,kerner}
 \begin{eqnarray}
 R_\pm(r,\theta_0)&=&\pm\int dr\sqrt{\frac{\Sigma(\theta_0)}{\Delta}}
 (\sqrt{-g^{tt}(\theta_0)}\omega+\frac{g^{t\phi}(\theta_0)}{\sqrt{-g^{tt}(\theta_0)}}j)\;\;,\nonumber\\
 &=&\pm\int \frac{dr}{\Delta}(\omega\sqrt{(r^2+a^2)^2-\Delta
 a^2\textrm{sin}^2\theta_0}-j\frac{a(r^2+a^2-\Delta)}{\sqrt{(r^2+a^2)^2-\Delta
 a^2\textrm{sin}^2\theta_0}})\;\;.\nonumber
 \end{eqnarray}
 The imaginary part of $R_+$ can be calculated using the above equation.
 Integrating the pole at the horizon leads to the result
 (see \cite{kerner,mitra} for a detailed similar process)
 \begin{eqnarray}
 \textrm{Im}R_\pm=\pm\frac{\pi(r^2+a^2)}{r_+-r_-}(\omega-j\Omega_H)\;\;,
 \end{eqnarray}
 where $\Omega_H=\frac{a}{r_+^2+a^2}$ is the angular velocity of
 event horizon. One can see that this result is independent of $\theta$.

 As discussed in the Hamilton-Jacobi method\cite{akhmedov,mitra}, one solution
 corresponds Dirac particles moving away from the outer event horizon and the
 other solution corresponds the particles moving toward the outer event
 horizon. The probabilities of crossing the outer horizon each way are
 respectively given by
 \begin{eqnarray}
 P_{out}&=&\textrm{exp}[-\frac{2}{\hbar}\textrm{Im}I]
 =\textrm{exp}[-\frac{2}{\hbar}(\textrm{Im}R_++\textrm{Im}K)]\;,\nonumber\\
 P_{in}&=&\textrm{exp}[-\frac{2}{\hbar}\textrm{Im}I]
 =\textrm{exp}[-\frac{2}{\hbar}(\textrm{Im}R_-+\textrm{Im}K)]\;.
 \end{eqnarray}
 To ensure that the probability is normalized, we should note
 that the probability of any incoming wave crossing
 the outer horizon is unity\cite{mitra}. So we get $\textrm{Im}K=-\textrm{Im}R_-$.
 Since $\textrm{Im}R_+=-\textrm{Im}R_-$ this implies that the
 probability of a Dirac particle tunneling from inside to outside the
 evnet horizon is given by
 \begin{eqnarray}
 \Gamma&=&\textrm{exp}[-\frac{4}{\hbar}\textrm{Im}R_+]\;\;,\nonumber\\
 &=&\textrm{exp}[\frac{4\pi(r_+^2+a^2)}{(r_+-r_-)}(\omega-j\Omega_H)]\;\;,
 \end{eqnarray}
 where in the last step we set $\hbar$ to unity.
 It should be noted that the higher terms about $\omega$ and $j$
 are neglected in our derivation and the expression (12) for tunneling
 probability implies the pure thermal radiation.

 From the tunneling probability, the fermionic
 spectrum of Hawking radiation of
 Dirac particles from Kerr black hole
 can be deduced following the standard arguments\cite{damour,sannan}
 \begin{eqnarray}
 N(\omega, j)=\frac{1}{e^{2\pi(\omega-j\Omega_H)/\kappa}+1}\;\;,
 \end{eqnarray}
 where $\kappa=\frac{(r_+-r_-)}{2(r_+^2+a^2)}$ is the surface
 gravity of event horizon.
 From the tunneling probability and radiant spectrum, Hawking temperature
 of Kerr black hole can be determined as
 \begin{eqnarray}
 T=\frac{\kappa}{2\pi}=\frac{\sqrt{M^2-a^2}}{4\pi M(M+\sqrt{M^2-a^2})}\;\;.
 \end{eqnarray}

 In summary, we have calculated the Dirac particles' Hawking
 radiation from Kerr black hole using the tunneling formalism.
 Starting with Dirac equation, we obtained the radiation spectrum
 and Hawking temperature of Kerr black hole by using
 the WKB approximation. The results
 coincide with the previous literatures\cite{zhangplb,angheben,kerner,jiang,kim}.

 \section*{ACKNOWLEDGEMENT}

 After completing this paper, we noticed that some related
 works were done by other authors. In \cite{KM}, charged
 Fermions tunnelling from Kerr-Newman black holes were
 investigated. In \cite{chen}, the authors considered
 Fermions tunnelling from the more
 general and complicated spacetime background.
 This work was supported by the National Natural Science Foundation
 of China and Cuiying Project of Lanzhou University.


\begin{thebibliography}{99}


 \bibitem{hawking}

 S.W.Hawking, Commun.Math.Phys.\textbf{43},199(1975).

 \bibitem{parikh}

 M.K.Parikh and F.Wilczek, Phys.Rev.Lett.\textbf{85},5042(2000).

 \bibitem{zhangjhep}

 J.Zhang and Z.Zhao, JHEP \textbf{10},055(2005).

 \bibitem{zhangplb}

 J.Zhang and Z.Zhao, Phys.Lett.B \textbf{618},14(2005);
 J.Zhang and Z.Zhao, Phys.Lett.B \textbf{638},110(2006).

 \bibitem{angheben}

 M.Angheben, M.Nadalini, L.Vanzo and S.zerbini, JHEP \textbf{05},
 014(2005).

 \bibitem{padmanabhan}

 K.Srinivasan and T.Padmanabhan, Phys.Rev.D \textbf{60},
 24007(1999); S.Shankaranarayanan, T.Padmanabhan and K.Srinivasan,
 Class.Quant.Grav. \textbf{19}, 2671(2002).

 \bibitem{kerner}

 R.Kerner and R.B.Mann, Phys.Rev.D \textbf{73}, 104010(2006).

 \bibitem{kernerarxiv}

 R.Kerner and R.B.Mann, Class.Quant.Grav. \textbf{25}, 095014(2008).

 \bibitem{liplb}

 R. Li, J. R. Ren, Phys. Lett. B, \textbf{661}, 370(2008).

 \bibitem{vanzo}

 R. Di Criscienzo and L. Vanzo, arXiv: 0803.0435.

 \bibitem{akhmedov}

 E.T.Akhmedov, V.Akhmedov and D.Singleton, Phys.Lett.B
 \textbf{642}, 124(2006).

 \bibitem{mitra}

 P.Mitra, Phys.Lett.B \textbf{648}, 240(2007).

 \bibitem{damour}

 T.Damour and R.Ruffini, Phys.Rew.D \textbf{14}, 332(1976).

 \bibitem{sannan}

 S.Sannan, Gen.Rel.Grav. \textbf{20}, 139(1988).

 \bibitem{jiang}

 Q.Jiang, S.Wu and X.Cai, Phys.Rev.D \textbf{73}, 064003(2006).

 \bibitem{kim}

 S.P.Kim, JHEP, \textbf{11}, 048(2007).

 \bibitem{KM}

 R. Kerner and R.B. Mann, arXiv:0803.2246.

 \bibitem{chen}

 D. Chen, Q. Jiang, S. Yang and X. Zu, arXiv:0803.3248;
 D. Chen, Q. Jiang and X. Zu, arXiv:0804.0131.

 \end{thebibliography}
 \end{document}